\documentclass[%
 reprint,
 amsmath,amssymb,
 aps, prl
]{revtex4-1}

\usepackage{graphicx}
\usepackage{dcolumn}
\usepackage{bm}
\usepackage{graphicx}
\usepackage{amsmath}
\usepackage{amsfonts}
\usepackage{amssymb}
\usepackage{xcolor}
\usepackage{soul}

\usepackage{footmisc}

\begin{document}

\preprint{APS/123-QED}

\title{Lateral Chirality-sorting Optical Spin Forces in Evanescent Fields}

\author{Amaury Hayat}%
\thanks{These authors contributed equally to this work.}
\affiliation{School of Engineering and Applied Sciences, Harvard University, Cambridge, Massachussetts 02138, USA}
\affiliation{\'Ecole Polytechnique, Palaiseau 91120, France}

\author{J. P. Balthasar M\"uller$^{*,}$}
\email{corresponding author: jpbm@seas.harvard.edu}
\affiliation{School of Engineering and Applied Sciences, Harvard University, Cambridge, Massachusetts 02138, USA}



\author{Federico Capasso}
\affiliation{School of Engineering and Applied Sciences, Harvard University, Cambridge, Massachusetts 02138, USA}


\date{\today}

\begin{abstract}
The transverse component of the spin angular momentum of evanescent waves gives rise to lateral optical forces on chiral particles, which have the unusual property of acting in a direction in which there is neither a field gradient nor wave propagation. As their direction and strength depends on the chiral polarizability of the particle, they act as chirality-sorting and may offer a mechanism for passive chirality spectroscopy. The absolute strength of the forces also substantially exceeds that of other recently predicted sideways optical forces, such that they may more readily offer an experimental confirmation of the phenomenon.
\begin{description}

\item[PACS numbers]
12345
\end{description}
\end{abstract}

\pacs{Valid PACS appear here}
\maketitle


\section{Introduction}

Chirality (from Greek \textit{$\chi \epsilon \iota \rho$ (kheir)}, 'hand') is a type of asymmetry where a geometric object can not be made to coincide with its mirror image through any proper spatial rotation \cite{kelvin}. While they share all properties other than their helicity, a chiral object and its mirror image differ in their interaction with a chiral environment, such as biological systems. The analysis of and separation by chirality of substances consequently represents a highly important challenge in research and industry, affecting especially pharmaceuticals and agrochemicals \cite{smith2009,pesticides,FDA92,caner2004}. While the sorting of substances by chirality normally has to be addressed through the introduction of a specific chiral resolving agent \cite{nguyen2006}, the manifestation of chirality in the electromagnetic response of materials has raised the question of passive sorting using optical forces \cite{canaguier2013, tkachenko2014,tkachenko2014a,cameron2014,smith2014, wang2014}.

In Ref. \cite{wang2014} Wang et. al. recently predicted an electromagnetic plane wave to exert a lateral optical force on a chiral particle above a reflective surface, which emerges as the particle interacts with the reflection of its scattered field. This highly unusual force acts in a direction in which there is neither wave propagation nor an intensity gradient, and deflects particles with opposite helicities towards opposite sides. Apart from it's fundamental interest, such a force may in theory be useful for all-optical enantiomer sorting with a single, unstructured beam. In Ref. \cite{bliokh2014}, Bliokh et. al. almost simultaneously predicted another lateral force that is exerted on non-chiral particles in an evanescent wave. This force is a consequence of the linear momentum of the wave that is associated with its spin angular momentum (SAM), which was first described by F. J. Belifante in the context of quantum field theory, and which vanishes for a propagating plane wave \cite{bliokhPRA2011,belifante}. 

In this letter we investigate a third mechanism, in which a force emerges as the spin angular momentum itself gives rise to a linear momentum transfer as it interacts with chiral materials. This effect enables the tailoring of chirality-sorting optical forces by engineering the local SAM density of optical fields, such that fields with transverse SAM can be used to achieve lateral forces. We calculate the forces emerging in an evanescent field with transverse SAM, and show that the magnitude of the lateral force substantially exceeds those of the previously predicted lateral forces.
\begin{figure}[h]
\includegraphics[width=3in]{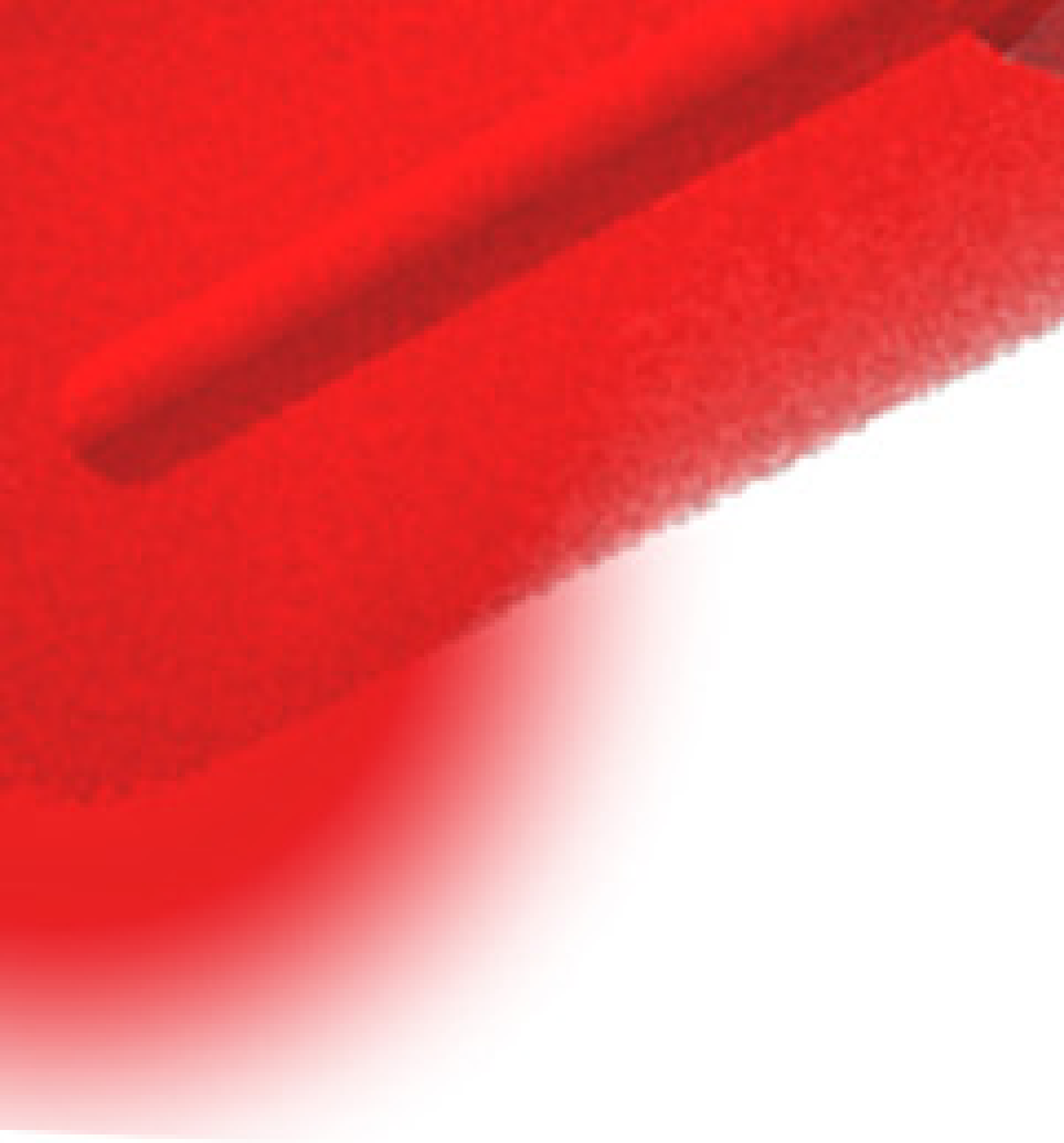}
\caption{\textbf{Chirality-dependent Lateral Forces in an Evanescent Field:} An evanescent field arises as light in a high index medium ($n_1$) is totally internally reflected at the interface with a low index medium ($n_2$) at an angle $\theta$ beyond the critical angle. Particles in an evanescent field with transverse spin angular momentum experience lateral forces depending on their chiral polarizability $\chi$, with particles with opposite helicities experiencing lateral forces in oppose directions.}
\label{Figure1}
\end{figure}
In the following section of this paper we provide a brief general discussion of the optical forces exerted by an electromagnetic wave on a small chiral particle and treat the specific case of an evanescent wave. We then calculate the lateral force on a nano sphere, and compare it to the strength of the lateral force that may arise due to the other two effects discussed by Wang et. al and Bliokh et. al \cite{wang2014,bliokh2014}.
\section{Optical Forces on Small Particles in an Evanescent Field}
The optical forces exerted on an object is rigorously calculated by first finding the distribution of electromagnetic fields and then integrating the Maxwell stress tensor over a surface enclosing the object. The scattering problem associated with finding the field distributions is tremendously simplified in the dipole approximation, which holds in the so-called Rayleigh limit that applies to particles much smaller than the wavelength. The dipole solution is sufficiently accurate for most molecular scattering problems and is furthermore the basis for efficient numerical methods that are used to find solutions for complex larger objects \cite{purcellDDA}. We consider the optical forces on a small particle in a source-free, lossless, non-dispersive and isotropic medium with relative electric permittivity $\epsilon$ and relative magnetic permeability $\mu$. In the dipole approximation, the time-averaged total optical force $\mathbf{F}$ exerted on the particle by a monochromatic electromagnetic wave is given by \cite{vesperinas2010}:
\begin{equation}
\label{force1}
\begin{split}\mathbf{F}=	\underbrace{	\frac{1}{2}\mathrm{Re}\left\{\mathbf{d}\left(\nabla\otimes\mathbf{E}^*\right)+\mathbf{m}\left(\nabla\otimes\mathbf{H^{*}}\right)\right\}	}_\mathbf{{F_{0}}} \\ -\underbrace{	\frac{k^4}{3}\mathrm{Re}\left\{\sqrt{\frac{1}{\mu\epsilon}}	\left(	\mathbf{d}\times{\mathbf{m}^{*}}	\right)	\right\}	}_\mathbf{{F_{\mathrm{int}}}},\end{split}
\end{equation}
where $\mathbf{E}$ and $\mathbf{H}$ are the electric and magnetic field vectors of the incident electromagnetic wave at the location of the particle; $\mathbf{d}$ and $\mathbf{m}$ are the electric and magnetic dipole moments of the particle; $k=n\omega/c$ is the wavenumber of the electromagnetic wave; $\omega$ is the frequency; $c$ is the speed of light in vacuum; and $n=\sqrt{\epsilon\mu}$ the refractive index of the medium. The fields are written in complex phasor notation throughout this paper, where the factor $\mathrm{exp}\left(-i\omega t\right)$ giving the time dependence is implied, and the superscript $^*$ denotes the complex conjugate. Vector quantities are indicated by bold letters, and all expressions are given in Gaussian units unless stated otherwise. $\mathbf{\hat{x}},\mathbf{\hat{y}}$ and $\mathbf{\hat{z}}$ are unit vectors along the corresponding coordinate axes. The symbol $\otimes$ denotes the dyadic product, so that the terms of the form $\mathbf{W}\left(\nabla\otimes\mathbf{V}\right)$ in eqn. \ref{force1} have elements $\left[\mathbf{W}\left(\nabla\otimes\mathbf{V}\right)\right]_i = \sum_j W_j \partial_i V_j$ for $i,j \in \{x,y,z\}$ \cite{vesperinas2010} and can be written as $\mathbf{W}\left(\nabla\otimes\mathbf{V}\right)=\left(\mathbf{W}\cdot\nabla\right)\mathbf{V} + \mathbf{W}\times\left(\nabla\times\mathbf{V}\right)$ in terms of more commonly used vector operators.
Equation \ref{force1} shows that the optical force has a component $\mathbf{F_{0}}$ corresponding to the force exerted by the incident field on the particle's electric and magnetic dipole moments, and a component $\mathbf{F_{\mathrm{int}}}$ that results from a direct interaction of the two dipole moments. In case of the evanescent field, it is the $\mathbf{F_{int}}$ term that gives rise to the lateral electromagnetic spin force.
Chirality manifests itself in the electromagnetic response of a material through a cross-coupling of the induced electric and magnetic polarizations, such that an electric field also gives rise to a magnetic polarization and vice versa. Indeed, the optical response of chiral materials may be modeled intuitively by a simple coupled oscillator model \cite{yin2014}. Assuming isotropic polarizabilities and no permanent dipole moment, the electric and magnetic dipole moments induced by fields incident on a chiral particle are $\bf{d}=\alpha_{\mathrm{e}}\bf{E}+\mathrm{i} \chi\bf{H}$  and $\bf{m}=\alpha_{\mathrm{m}}\bf{H}-\mathrm{i} \chi\bf{E}$, where $\alpha_{\mathrm{e}}$ and $\alpha_{\mathrm{m}}$ are the electric and magnetic polarizabilities \cite{wang2014}. The chiral polarizability of the particle $\chi$ captures the chiral nature of the dipole, such that setting $\chi=0$ recovers the case of an achiral dipole. It is worth mentioning that, while the chiral polarizability is an inherently dynamic quantity, the dynamic dipole polarizabilities  $\alpha_\mathrm{e}$ and $\alpha_\mathrm{m}$ differ from the more frequently listed corresponding static polarizabilities $\alpha_\mathrm{e}^{(0)}$ and $\alpha_\mathrm{m}^{(0)}$ by a radiation correction first derived by Draine as  $\alpha_{\mathrm{e}} = \left[1-\left(i2k^3/3\epsilon\right)\alpha_{\mathrm{e}}^{(0)} \right]^{-1}\alpha_{\mathrm{e}}^{(0)}$ and $\alpha_{\mathrm{m}} = \left[1-\left(i2k^3/3\mu\right)\alpha_{\mathrm{m}}^{(0)} \right]^{-1}\alpha_{\mathrm{m}}^{(0)}$ \cite{draine}. While the static polarizability is sufficiently accurate in the particular case of the small gold sphere considered in ref. \cite{bliokh2014}, omission of the radiation correction can in general lead to a substantial underestimation of the optical forces \cite{supplementary}. Inserting the expressions for the electric and magnetic dipole moments into equation \ref{force1} gives the force on a chiral dipole in an electromagnetic field:
\begin{widetext}
\begin{equation}
\label{f1}
\mathbf{F_{\mathrm{0}}}=\underbrace{\frac{1}{g}\left( \epsilon^{-1}\mathrm{Re}\left\{\alpha_{\mathrm{e}}\right\}\nabla {u}_{\mathrm{e}}+  \mu^{-1} \mathrm{Re}\left\{\alpha_{\mathrm{m}}\right\}\nabla {u}_{\mathrm{m}}\right)-\frac{\omega}{g} \mathrm{Re}\left\{\chi\right\}\nabla h}_{\mathrm{Gradient\ Force}}+\underbrace{\frac{2\omega}{g}(\mu\mathrm{Im}\left\{\alpha_{\mathrm{e}}\right\}\mathbf{p_{\mathrm{e}}^{o}}+\epsilon\mathrm{Im}\left\{\alpha_{\mathrm{m}}\right\}\mathbf{p_{\mathrm{m}}^{o}}) -\frac{c}{g}\mathrm{Im}\left\{\chi\right\}[\nabla\times\mathbf{p}-2k^{2}\mathbf{s}]}_{\mathrm{Radiation\ Pressure}}
\end{equation}
\begin{equation}
\label{f2}
\mathbf{F_{\mathrm{int}}}=-\frac{2\omega}{g}\frac{k^{3}}{3}(\mathrm{Re}\left\{\alpha_{\mathrm{e}} \alpha_{\mathrm{m}}^{*}\right\}\mathbf{p}-\mathrm{Im}\left\{\alpha_{\mathrm{e}} \alpha_{\mathrm{m}}^{*}\right\}\mathbf{{p''}}+\arrowvert\chi\arrowvert^{2}\mathbf{p})-\frac{2\omega}{g}\frac{2k^{4}}{3n}( \epsilon\mathrm{Re}\left\{\chi\alpha_{\mathrm{m}}^{*}\right\}\mathbf{s_{\mathrm{m}}}+ \mu\mathrm{Re}\left\{\chi\alpha_{\mathrm{e}}^{*}\right\}\mathbf{s_{\mathrm{e}}}).
\end{equation}
\end{widetext}
Here we marked the terms corresponding to the familiar gradient force and radiation pressure, both of which are modified in the presence of chirality. The final term of eqn. \ref{f2} shows that the spin momentum density of the field gives rise to a linear momentum transfer on chiral particles, which occurs in opposite directions for opposite signs of $\chi$, i.e. opposite helicities. The prefactor $g=\left(4\pi\right)^{-1}$ arises from the usage of Gaussian units, ${u}_\mathrm{e} = (g\epsilon/4)|\mathbf{E}|^2$ and ${u}_\mathrm{m} = (g\mu/4)|\mathbf{H}|^2$ are the energy densities of the electric and the magnetic field, and $h = (g/2\omega)\mathrm{Im} \left\{\mathbf{E}\cdot\mathbf{H^{*}}\right\}$ is the time-averaged optical helicity density \cite{bliokh2014,bliokhPRA2011}. The time-averaged Poynting momentum density $\mathbf{p}=(g/2c)\mathrm{Re}\left\{\mathbf{E}\times\mathbf{H^{*}}\right\}$ can be decomposed as $\mathbf{p} = \mathbf{p^o}+\mathbf{p^s}$ into a component related to the orbital angular momentum,  $\mathbf{p^o}$, and the Belifante momentum $\mathbf{p^s}$ related to the spin momentum density $\mathbf{s}$. These quantities can be individually further separated into their electric and magnetic components, in particular $\mathbf{p^o} = \mathbf{p^o_e}+\mathbf{p^o_m}$ and $\mathbf{s} = \mathbf{s_e}+\mathbf{s_m}$ (the supplementary material \cite{supplementary} contains a full list of the definitions of the field quantities, which are discussed in detail in ref. \cite{bliokh2014}). Invoking a fluid mechanics analogy, the term $\nabla\times\mathbf{p}$ may be interpreted as the vorticity of the Poynting vector flow. $p'' = (g/2c)\mathrm{Im}\left\{\mathbf{E}\times\mathbf{H^{*}}\right\}$ is the imaginary Poynting momentum \cite{bliokh2014,jackson}.  For a monochromatic wave, the momentum and spin densities are \cite{bliokh2014}:
\begin{eqnarray}
\label{quant1}
\mathbf{p^{s}}=-\frac{g}{8\omega}\nabla\times\left[\left(i\mu \right)^{-1}\mathbf{E}\times\mathbf{E^{*}}+\left(i\epsilon\right)^{-1}\mathbf{H}\times\mathbf{H^{*}}\right] \\
\mathbf{p^{o}}=-\frac{g}{4\omega}\mathrm{Im}\left\{\mu^{-1}\mathbf{E}\left(\nabla\otimes\mathbf{E^*}\right)+\epsilon^{-1}\mathbf{H}\left(\nabla\otimes\mathbf{H^*}\right)\right\}\\
\mathbf{s}=-\frac{g}{4\omega}\left((\mu i)^{-1}\mathbf{E}\times\mathbf{E^{*}}+(\epsilon i)^{-1}\mathbf{H}\times\mathbf{H^{*}}\right),\label{quant2}
\end{eqnarray}
where the magnetic and electric contributions correspond to the terms proportional to the electric and magnetic fields. Evaluating equations \ref{f1}-\ref{f2} for a plane wave  $\mathbf{E}=\sqrt{\mu}\mathbf{E_0}\mathrm{exp}(ikz)$ propagating in the $\mathbf{\hat{z}}$-direction, shows that the forces are entirely in the direction of propagation: 
\begin{eqnarray}
\begin{array}{rr}
\mathbf{F_{0}} = & \left[\left(\mathrm{Im}\left\{\mu\alpha_\mathrm{e}\right\}+\mathrm{Im}\left\{\epsilon\alpha_\mathrm{m}\right\}\right)/2\right. \\ 
& +\left.\mathrm{Im}\left\{\chi\right\}n\sigma\right]Ik\ \mathbf{\hat{z}}
\end{array}\\
\label{planewave}
\begin{array}{rr}
\mathbf{F_{int}} & = -\left[ \left(\epsilon\mathrm{Re}\left\{\chi\alpha_\mathrm{m}^*\right\} +\mu\mathrm{Re}\left\{\chi\alpha_\mathrm{e}^*\right\}		\right)\sigma/n		\right. \\
& + \left.  \mathrm{Re}\left\{\alpha_\mathrm{e}\alpha_\mathrm{m}^*\right\}  + |\chi|^2 \right]Ik^4/3\ \mathbf{\hat{z}}
\end{array}
\end{eqnarray}
Here $I = n^{-1} {|\mathbf{E}\times\mathbf{H^{*}}|}=|\mathbf{E_0}|^2$ is the intensity. The factor $\sigma = -2\mathrm{Im}\left\{\left(\mathbf{E}\cdot\mathbf{\hat{x}}\right)\left(\mathbf{E^*}\cdot\mathbf{\hat{y}}\right)\right\}/\left(|\mathbf{E}\cdot\mathbf{\hat{x}}|^2+|\mathbf{E}\cdot\mathbf{\hat{y}}|^2\right)$ is a measure for the degree of circular polarization (ellipticity) of the wave in the $(\mathrm{x},\mathrm{y})$-plane, such that $\sigma = +1$ for right circular polarization and $\sigma = -1$ for left circular polarization. In the calculations that follow, we will furthermore use the parameters $\tau=\left(|\mathbf{E}\cdot\mathbf{\hat{x}}|^2-|\mathbf{E}\cdot\mathbf{\hat{y}}|^2\right)/\left(|\mathbf{E}\cdot\mathbf{\hat{x}}|^2+|\mathbf{E}\cdot\mathbf{\hat{y}}|^2\right)$ and $\xi=2\mathrm{Re}\left\{(\mathbf{E}\cdot\mathbf{\hat{x}})(\mathbf{E^*}\cdot\mathbf{\hat{y}})\right\}/\left(|\mathbf{E}\cdot\mathbf{\hat{x}}|^2+|\mathbf{E}\cdot\mathbf{\hat{y}}|^2\right)$ as measures for the linear polarization in the $(\mathrm{x},\mathrm{y})$-plane, such that $\tau=\{+1,-1\}$ correspond to linear polarization along $\{x,y\}$ and $\xi=\{+1,-1\}$ correspond to linear polarization at $\{+45^{\circ},-45^\circ\}$ with respect to the $x$-axis. The helicity-dependent change of the radiation pressure due to the chiral polarizability $\chi$ in equation \ref{planewave} has previously been used for sorting highly chiral liquid crystal droplets in optical lattices \cite{tkachenko2014}. We now may consider an evanescent wave created by the total internal reflection of light at an interface in the (z,y)-plane, which has the form $\mathbf{E}=\sqrt{\mu}\mathbf{E_0}\mathrm{exp}(-\kappa x)\mathrm{exp}(ik_\mathrm{z}z)$. If the evanescent field is propagating in a medium with index $n_2=\sqrt{\epsilon\mu}$ and was created through the total internal reflection of a beam in a medium with index $n_1$ incident on the interface with angle $\theta$ in the (x,z)-plane (Fig. \ref{Figure1}), then the wave vector components are given by $k_z = \left(n_1/ n_2\right)\mathrm{sin}\left(\theta\right)k$, $k_x = \sqrt{k^2 - k_z^2}$ and $\kappa=\sqrt{k_z^2 - k^2}=-ik_x$, where $k=n_2 \omega/c$. The expressions for all quantities relevant for the calculation of the forces with eqns. \ref{f1} and \ref{f2} are listed in the supplementary information \cite{supplementary}. Notably, evanescent fields have longitudinally polarized field components \cite{supplementary}, which give rise to transverse spin angular momentum \cite{bliokhspin,bliokh2014}. In case of a p-polarized wave ($\tau = 1$) it is the electric field that is elliptically polarized in the (x,z)-plane, and in case of a s-polarized wave ($\tau = -1$) it is the magnetic field. Correspondingly, the out of plane electric and magnetic spin-components are given by $(\mathbf{\hat{y}}\cdot\mathbf{s_\mathrm{e}}) = (1+\tau)\frac{g}{4\omega} \frac{\kappa}{k_{z}}   I_\mathrm{e}$ and $(\mathbf{\hat{y}}\cdot\mathbf{s_\mathrm{m}}) = (1-\tau)\frac{g}{4\omega} \frac{\kappa}{k_{z}}   I_\mathrm{e}$ and the full expressions for the forces on a chiral dipole in an evanescent field are:
\begin{widetext}
\begin{equation}\begin{split}
\mathbf{F_{\mathrm{0}}}=\kappa I_{\mathrm{e}}[\mathrm{Re}\{\chi\}n_{\mathrm{2}}\sigma-\frac{1}{2}(\mathrm{Re}\{\mu\alpha_{\mathrm{e}}\}(1+\tau\frac{\kappa^{2}}{k_{\mathrm{z}}^{2}})+\mathrm{Re}\{\epsilon\alpha_{\mathrm{m}}\}(1-\tau\frac{\kappa^{2}}{k_{\mathrm{z}}^{2}}))]\mathbf{\hat{x}}\\
\frac{I_{\mathrm{e}}}{2}k_{\mathrm{z}}[\mathrm{Im}\{\mu\alpha_{\mathrm{e}}\}(1+\tau\frac{\kappa^{2}}{k_{\mathrm{z}}^{2}})+\mathrm{Im}\{\epsilon\alpha_{\mathrm{m}}\}(1-\tau\frac{\kappa^{2}}{k_{\mathrm{z}}^{2}})+2\mathrm{Im}\{\chi\}\sigma n_{\mathrm{2}}]\mathbf{\hat{z}}
\end{split}\end{equation}
\begin{equation}
\begin{split}
\mathbf{F_{\mathrm{int}}}=-I_{\mathrm{e}} \frac{k^{4}}{3} \frac{\kappa}{k_{z}}[\mathrm{Re}\{\alpha_{\mathrm{e}}\alpha^{*}_{\mathrm{m}}\}\sigma +\mathrm{Im}\{\alpha_{\mathrm{e}}^{*}\alpha_{\mathrm{m}}\}\xi+\sigma\arrowvert\chi\arrowvert^{2}+\frac{\epsilon}{n_{2}}(1-\tau)\mathrm{Re}\{\chi\alpha_{\mathrm{m}}^{*}\}+\frac{\mu}{n_{2}}(1+\tau)\mathrm{Re}\{\chi\alpha_{\mathrm{e}}^{*}\}]\mathbf{\hat{y}}\\
-I_{\mathrm{e}}\frac{k^{4}}{3}\frac{k}{k_{\mathrm{z}}}[\mathrm{Re}\{\alpha_{\mathrm{e}}\alpha^{*}_{\mathrm{m}}\}+\arrowvert\chi\arrowvert^{2}+\frac{\sigma}{n_{2}}(\epsilon \mathrm{Re}\{\chi\alpha_{\mathrm{m}}^{*}\}+\mu \mathrm{Re}\{\chi\alpha_{\mathrm{e}}^{*}\})]\bf{\hat{z}}\\
-I_{\mathrm{e}}\frac{k^{4}}{3}\frac{\kappa k}{k_{\mathrm{z}}^{2}}(\frac{\xi}{n_{\mathrm{2}}}(\mathrm{Re}\{\mu\chi\alpha_{\mathrm{e}}^{*}\}-\mathrm{Re}\{\epsilon\chi\alpha_{\mathrm{m}}^{*}\})+\mathrm{Im}\{\alpha_{\mathrm{e}}^{*}\alpha_{\mathrm{m}}\}\tau)\mathbf{\hat{x}}\label{chirlat}\end{split}
\end{equation}
\end{widetext}
with $I_{\mathrm{e}} = \frac{|\mathbf{E_0}|^2}{1+\tau (\kappa/k_{\mathrm{z}})^{2}} \mathrm{exp}(-2\kappa x)$. The lateral forces experienced by the dipole are given by the first line of eqn. \ref{chirlat}, with all chirality-dependent terms proportional to $\chi$. Unfortunately the chirality-dependent lateral forces caused by $\nabla\times\mathbf{p}$ and $\mathbf{s}$ in for the $\mathbf{F_0}$ force exactly cancel each other \cite{bliokhmagneto}. The $y$-component of the Poynting momentum density $\mathbf{p}$ arises entirely from the Belifante spin momentum density $\mathbf{p^s}$ for polarizations with non-vanishing $\sigma$, which also gives rise to a helicity-independent lateral force term on chiral particles. An analogous $y$-term arises as part of the imaginary Poynting momentum density $\mathbf{p''}$ for polarizations with non-vanishing $\xi$. The intensity and polarization of the evanescent wave depend on the complex transmission coefficients for the totally internally reflected beam. 

\begin{figure}[h]
\includegraphics[width=3in]{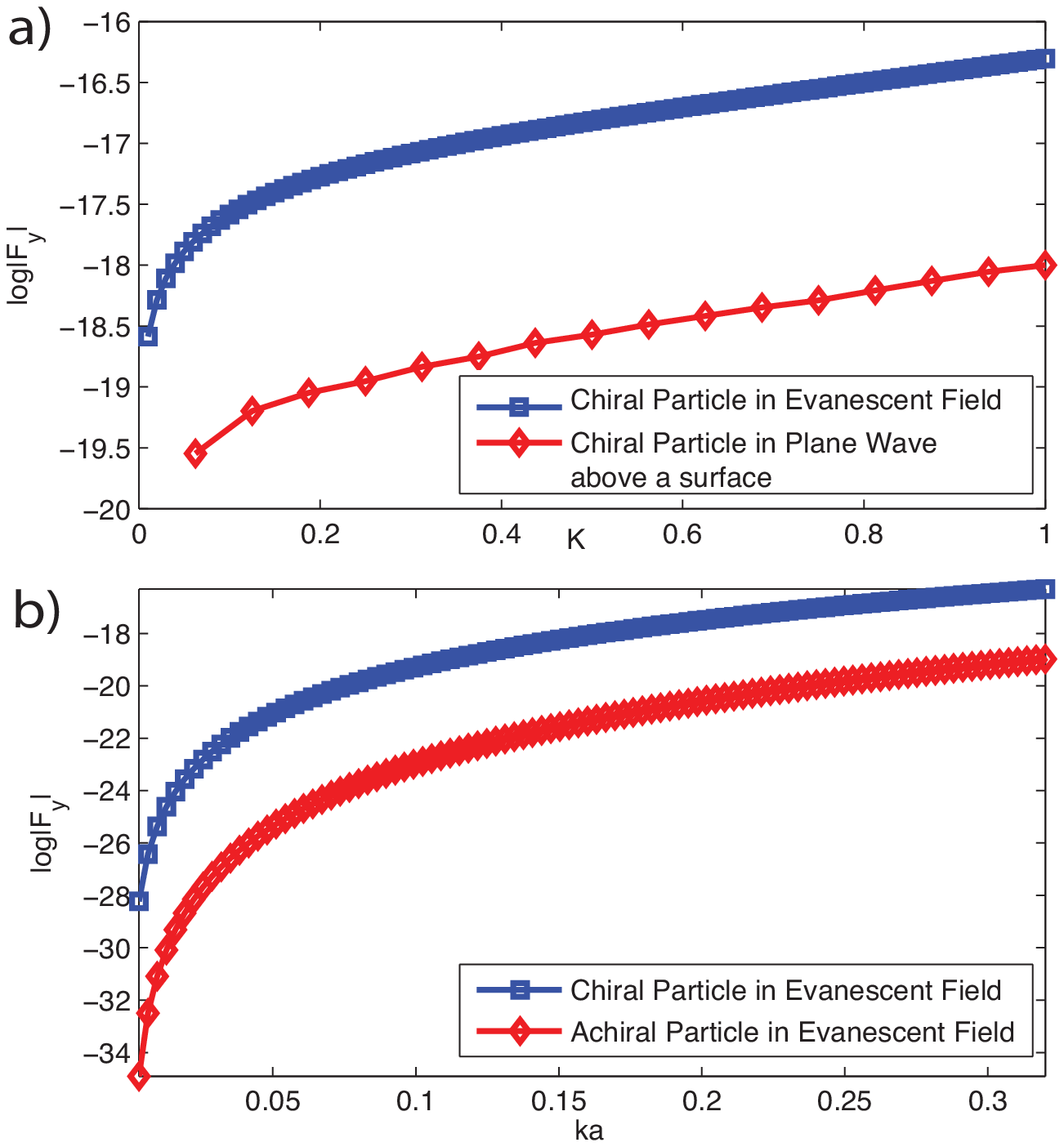}
\caption{\textbf{Lateral Optical Forces:} The previously predicted lateral optical forces in comparison to the lateral force on a chiral particle in an evanescent field. \textbf{a)} The lateral optical force on a chiral nano sphere in a plane wave above a reflective surface according to Wang et. al. \cite{wang2014} and due to an evanescent field in N/(mW/$\mathrm{\mu m^2}$) as a function of the chirality parameter $K$ \cite{Lakhtakia,supplementary}. The sphere is non-magnetic, situated 60nm above the surface and has a dielectric constant of $\epsilon_\mathrm{s} =2$ and a radius of 30nm. In case of the force in the evanescent field, the intensity refers to the intensity of a beam that is totally internally reflected at a flint glass/air interface at an angle of $\theta = 36.4^\circ$. \textbf{b)} The lateral force on an achiral sphere  ($K=0$) due to the Belifante spin momentum density according to Bliokh et. al. and on an equivalent chiral sphere ($K=1$) as a function of $ka$, where $a$ is the radius of the sphere and $k = n_{2}\omega/c$ is the wavenumber of the evanescent field. In both cases the sphere is non-magnetic and situated at a height of 30nm, has dielectric constant $\epsilon_\mathrm{s} =2$ and experiences the evanescent field generated at a flint glass/air interface at an incident angle of $\theta = 36.4^\circ$.}
\label{SOFFig2}
\end{figure}

To illustrate characteristic magnitudes of the force, the strength of the lateral force acting on a spherical nanoparticle is shown in Fig \ref{SOFFig2} as a function of chirality and particle radius in direct comparison with the lateral forces that arise due to the mechanisms described by Wang et. al \cite{wang2014} and Bliokh et. al \cite{bliokh2014}. In each case, the lateral force emerging through the interaction of the transverse SAM with the chiral particle is stronger by about 1.75 and four orders of magnitude over the considered range. Here material chirality is parametrized with a chirality parameter $K\in [-1,1]$  as in ref. \cite{wang2014}, and the derivation of the expression of the force on such a sphere is listed in the supplementary information \cite{supplementary}, as well as a discussion of the force on helicene molecules and gold nanohelices, which represent other types of chiral particles frequently considered in the literature. 

\section{Conclusion}
We predict lateral forces on materials with chiral optical response in evanescent fields, which push particles with opposite helicities in opposite directions with strength dependent on the chiral polarizability. The forces result from the interaction of the evanescent field's transverse optical SAM density of the wave with the chiral electromagnetic response of the particle, and are particularly strong in comparison to previously predicted lateral optical forces. Transverse SAM may arise whenever light is laterally confined, so that the effect described in this paper represents a natural choice for the optical sorting by material chirality in an integrated system. The effect can be produced by a single beam, which significantly aids alignment and avoids standing wave patterns that limit the separation of enantiomers to the width of interference fringes. However, the use of multiple beams may enable the cancellation of the longitudinal force component or, given the possibility of generating transverse spin, even the generation of spin-optical lateral forces in free space beams \cite{bliokh2014b}. While we limited the discussion to the simplest case of an evanescent field created by the total internal reflection of light at a single interface, there is significant potential for improving the strength of the force by further engineering the light field. Natural choices for this are the intense and inherently evanescent fields of plasmonic excitations, or optical waveguides designed to have regions with high field enhancement, such as slot waveguides \cite{slot}. We furthermore limited the discussion to the Rayleigh limit, where compact closed form expressions exist, though the optical manipulation of objects smaller than a few 100 nm in liquid suspension is in general very challenging due to thermal agitation. Other than by tailoring the light fields, this may in particular be overcome in a low pressure environment. Larger particles may take advantage of resonances (such as Mie resonances) and are bound to experience much stronger forces.
\section{Acknowledgements}
This research was supported by the Airforce Office of Scientific Research (AFOSR) under Grant no. FA9550-12-1-0289. The authors thank X. Yin for helpful discussions.

\begin{widetext}

 \end{widetext}
 
\end{document}


\title{Supplementary Material}

\maketitle

\section{Definition of the Field Quantities}

\begin{center}
\begin{tabular}{|c|p{5cm}|}

\hline
Definition of the quantity & Description\\
\hline
\hline
$\mathbf{p}=(g/2c)\mathrm{Re}\left\{\mathbf{E}\times\mathbf{H^{*}}\right\}$ & Poynting momentum density\\
\hline
$\mathbf{p''}=(g/2c)\mathrm{Im}\left\{\mathbf{E}\times\mathbf{H^{*}}\right\}$ & Value of the imaginary adjoint to the Poynting momentum density\\
\hline
\hline
$\mathbf{p^{o}}=-\frac{g}{4\omega}\mathrm{Im}\left\{\mu^{-1}\mathbf{E}\left(\nabla\otimes\mathbf{E^*}\right)+\epsilon^{-1}\mathbf{H}\left(\nabla\otimes\mathbf{H^*}\right)\right\}$ & Orbital (or canonical) poynting momentum density\\
\hline
$\mathbf{p^{o}_{\mathrm{e}}}=-\frac{g}{4\omega}\mathrm{Im}\left\{\mu^{-1}\mathbf{E}\left(\nabla\otimes\mathbf{E^*}\right)\right\}$ & Electric contribution to the orbital poynting momentum density\\
\hline
$\mathbf{p^{o}_{\mathrm{m}}}=-\frac{g}{4\omega}\mathrm{Im}\left\{\epsilon^{-1}\mathbf{H}\left(\nabla\otimes\mathbf{H^*}\right)\right\}$ & Magnetic contribution to the orbital poynting momentum density\\
\hline
\hline
$\mathbf{p^{s}}=-\frac{g}{8\omega}\nabla\times\left[\left(i\mu \right)^{-1}\mathbf{E}\times\mathbf{E^{*}}+\left(i\epsilon\right)^{-1}\mathbf{H}\times\mathbf{H^{*}}\right]$  & Spin part of the Poynting momentum density (Belinfante Spin Momentum)\\
\hline
$\mathbf{p^{s}_{\mathrm{e}}}=-\frac{g}{8\omega}\nabla\times\left[\left(i\mu \right)^{-1}\mathbf{E}\times\mathbf{E^{*}}\right]$ & Electric contribution to the spin momentum density\\
\hline
$\mathbf{p^{s}_{\mathrm{m}}}=-\frac{g}{8\omega}\nabla\times\left[\left(i\epsilon\right)^{-1}\mathbf{H}\times\mathbf{H^{*}}\right]$ & Magnetic contribution to the spin momentum density\\
\hline
\hline
$\mathbf{s}=-\frac{g}{4\omega}\left((\mu i)^{-1}\mathbf{E}\times\mathbf{E^{*}}+(\epsilon i)^{-1}\mathbf{H}\times\mathbf{H^{*}}\right)$ & Spin Angular Momentum (SAM) density\\
\hline
$\mathbf{s_{\mathrm{e}}}=-\frac{g}{4\omega}\left((\mu i)^{-1}\mathbf{E}\times\mathbf{E^{*}}\right)$ & Electric contribution to the SAM density\\
\hline
$\mathbf{s_{\mathrm{m}}}=-\frac{g}{4\omega}\left((\epsilon i)^{-1}\mathbf{H}\times\mathbf{H^{*}}\right)$ & Magnetic contribution to the Spin Angular Momentum \\
\hline
\hline
$h= (g/2\omega)\mathrm{Im} \left\{\mathbf{E}\cdot\mathbf{H^{*}}\right\}$ & Helicity \\
\hline
$\nabla\times\mathbf{p}=\frac{g}{2c}\nabla\times\mathrm{Re}\left\{\mathbf{E}\times\mathbf{H^{*}}\right\}$ & Vorticity of the photon flow \\
\hline
\hline
$u_{\mathrm{e}}=\frac{g}{4}\epsilon\arrowvert \mathbf{E}\arrowvert^{2}$ & Electric contribution to the energy density \\
\hline
$u_{\mathrm{m}}=\frac{g}{4}\mu\arrowvert \mathbf{H}\arrowvert^{2}$ & Magnetic contribution to the energy density\\
\hline
$I=n^{-1}|\mathbf{E}\times\mathbf{H}^{*}|$ & Intensity\\
\hline
\hline
\end{tabular}
\caption{\textbf{Supplementary Table 1:} Definitions of the field quantities relevant for the emergence of the optical forces on small chiral particles. A derivation and discussion of many quantities can be found in Bliokh et. al (2014) \cite{bliokh2014}
\end{center}

\newpage
\section{Field Quantities of a Propagating Wave and an Evanescent Wave}

\begin{center}
\begin{tabular}{|c|c|c|p{5cm}|}

\hline
Symbol & For a propagating wave
& For an evanescent field
  & Quantity\\
& $\mathbf{E}=\sqrt{\mu}\mathbf{E_0}\mathrm{exp}(ikz)$& 
$\mathbf{E}=\sqrt{\mu}\mathbf{E_0}\mathrm{exp}(-\kappa x)\mathrm{exp}(ik_\mathrm{z}z)$& \\
\hline
\hline
& $I=|\mathbf{E_{0}}|^{2}$ & $I_{\mathrm{e}}=\frac{|\mathbf{E_{0}}|^{2}}{1+\tau(\kappa/k_{\mathrm{z}})^{2}}\exp(-2\kappa x)$ & \\
\hline
\hline
$\mathbf{p}$ &$\frac{g}{2\omega}Ik_{\mathrm{z}}\mathbf{\hat{z}}$ &$\frac{g}{2\omega}I_{\mathrm{e}}(\sigma\frac{\kappa k}{k_{\mathrm{z}}}\mathbf{\hat{y}}+\frac{k^{2}}{k_{\mathrm{z}}}\mathbf{\hat{z}})$& Poynting momentum density\\
\hline
$\mathbf{p''}$ & 0 & $\frac{g}{2\omega}I_{\mathrm{e}}(\tau\frac{\kappa k^{2}}{k_{z}^{2}}\mathbf{\hat{x}}+\xi\frac{\kappa k}{k_{z}}\mathbf{\hat{y}})$ & Value of the imaginary adjoint to the Poynting momentum density\\
\hline
\hline
$\mathbf{p^{o}}$ & $\frac{g}{2\omega} Ik_{z}\mathbf{\hat{z}}$ &$\frac{g}{2\omega} I_{\mathrm{e}}k_{z}\mathbf{\hat{z}}$ & Orbital (or canonical) poynting momentum density\\
\hline
$\mathbf{p^{o}_{\mathrm{e}}}$ & $\frac{g}{4\omega} Ik_{z}\mathbf{\hat{z}}$ &$\frac{g}{4\omega} I_{\mathrm{e}}(1+\frac{\kappa^{2}}{k_{z}^{2}})k_{z}\mathbf{\hat{z}}$ & Electric contribution to the orbital poynting momentum density\\
\hline
$\mathbf{p^{o}_{\mathrm{m}}}$ & $\frac{g}{4\omega} Ik_{z}\mathbf{\hat{z}}$ & $\frac{g}{4\omega} I_{\mathrm{e}}(1-\frac{\kappa^{2}}{k_{z}^{2}})k_{z}\mathbf{\hat{z}}$ & Magnetic contribution to the orbital poynting momentum density\\
\hline
\hline
$\mathbf{p^{s}}$ & 0 & $\frac{g}{2\omega} I_{\mathrm{e}}(-\frac{\kappa^{2}}{k_{z}^{2}}\mathbf{\hat{z}}+\sigma\frac{\kappa k}{k_{z}}\mathbf{\hat{y}})$ & Spin part of the Poynting momentum density (Belinfante's momentum \cite{Belinfante})\\
\hline
$\mathbf{p^{s}_{\mathrm{e}}}$ & 0 & $\frac{g}{4\omega} I_{\mathrm{e}}(-(1+\tau)\frac{\kappa^{2}}{k_{z}^{2}}\mathbf{\hat{z}}+\sigma\frac{\kappa k}{k_{z}}\mathbf{\hat{y}})$ & Electric contribution to the spin momentum density\\
\hline
$\mathbf{p^{s}_{\mathrm{m}}}$ & 0 & $ \frac{g}{4\omega} I_{\mathrm{e}}(-(1-\tau)\frac{\kappa^{2}}{k_{z}^{2}}\mathbf{\hat{z}}+\sigma\frac{\kappa k}{k_{z}}\mathbf{\hat{y}})$ & Magnetic contribution to the spin momentum density\\
\hline
\hline
$\mathbf{s}$ & $\frac{g}{2\omega} I\sigma\mathbf{\hat{z}}$  & $\frac{g}{2\omega} I_{\mathrm{e}}(\sigma\frac{k}{k_{z}}\mathbf{\hat{z}}+\frac{\kappa}{k_{z}}\mathbf{\hat{y}})$ & Spin Angular Momentum (SAM) density\\
\hline
$\mathbf{s_{\mathrm{e}}}$ & $\frac{g}{4\omega} I\sigma\mathbf{\hat{z}}$  & $\frac{g}{4\omega} I_{\mathrm{e}}(\sigma\frac{k}{k_{z}}\mathbf{\hat{z}}+(1+\tau)\frac{\kappa}{k_{z}}\mathbf{\hat{y}}+\xi\frac{\kappa k}{k_{z}^2}\mathbf{\hat{x}})$ & Electric contribution to the SAM density\\
\hline
$\mathbf{s_{\mathrm{m}}}$ & $\frac{g}{4\omega} I\sigma\mathbf{\hat{z}}$ & $\frac{g}{4\omega} I_{\mathrm{e}}(\sigma\frac{k}{k_{z}}\mathbf{\hat{z}}+(1-\tau)\frac{\kappa}{k_{z}}\mathbf{\hat{y}}-\xi\frac{\kappa k}{k_{z}^2}\mathbf{\hat{x}})$& Magnetic contribution to the Spin Angular Momentum \\
\hline
\hline
$h$ & $\frac{g}{2\omega}In\sigma$ & $\frac{g}{2\omega}I_{\mathrm{e}}n_{2}\sigma$ & Helicity \\
\hline
$\nabla\times\mathbf{p}$ & 0 & $\frac{g}{2\omega}I_{\mathrm{e}}(2\frac{\kappa k^{2}}{k_{z}}\mathbf{\hat{y}}-2\sigma\frac{\kappa^{2} k}{k_{\mathrm{z}}}\mathbf{\hat{z}})$ & Vorticity of the photon flow \\
\hline
\hline
$u_{\mathrm{e}}$ & $\frac{g n^{2}}{4}I$ & $\frac{g n_{2}^{2}}{4}I_{\mathrm{e}}(1+\tau\frac{\kappa^{2}}{k_{\mathrm{z}}^{2}})$ & Electric contribution to energy density \\
\hline
$u_{\mathrm{m}}$ &$\frac{g n^{2}}{4}I$ &$\frac{g n_{2}^{2}}{4}I_{\mathrm{e}}(1-\tau\frac{\kappa^{2}}{k_{\mathrm{z}}^{2}})$ & Magnetic contribution to energy density\\
\hline
\end{tabular}
\caption{\textbf{Supplementary Table 2} Summary of the physical quantities involved in the force for a propagating plane wave and for an evanescent field. The magnetic and electric contributions to the orbital and spin angular momenta are listed explicitly.}
\end{center}

\section{Forces depending on Dynamic and Static Polarizabilities}

\subsection{Draine's Correction}
\begin{figure}[h]
\includegraphics[width=6in]{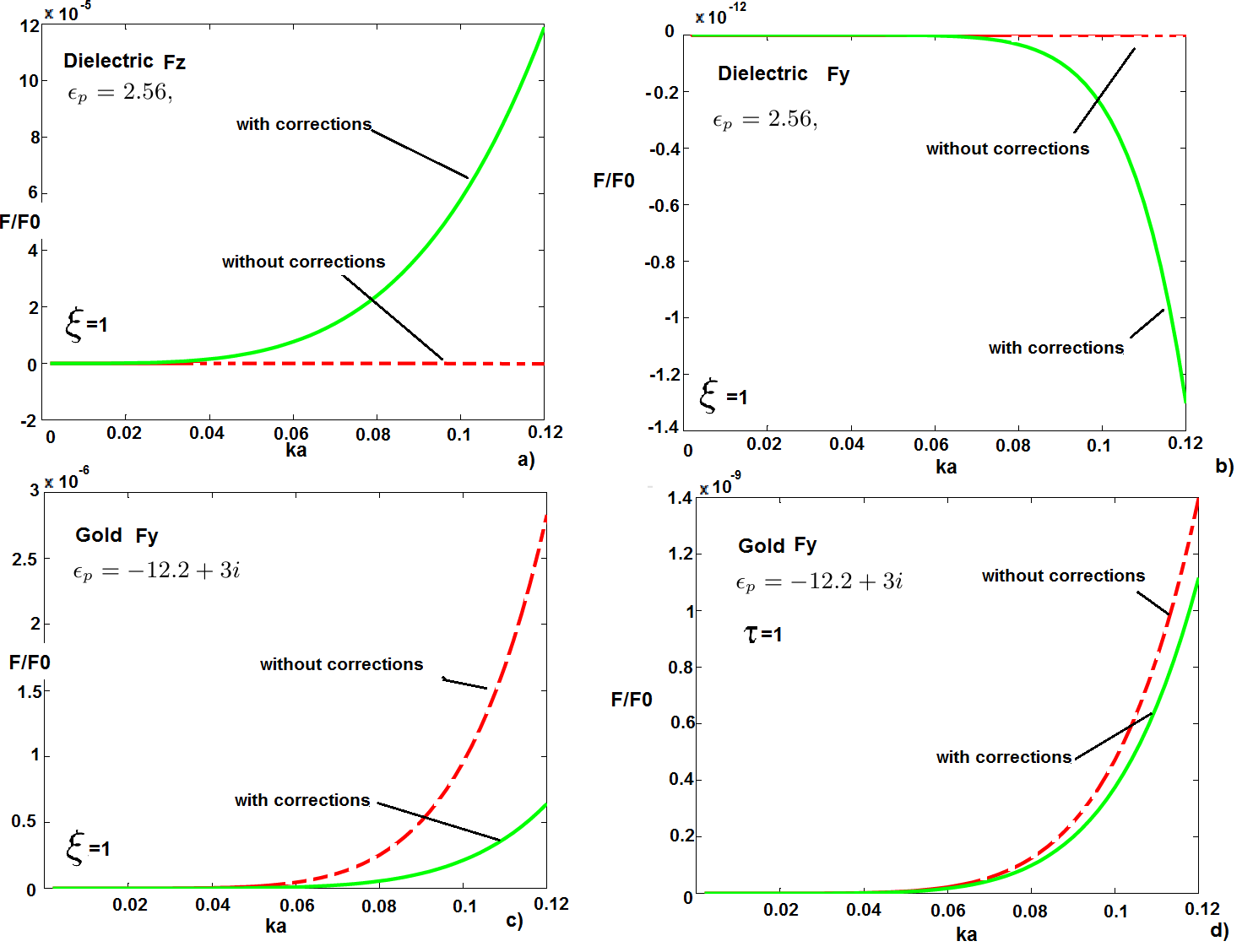}

\caption{\textbf{Impact of Draine's correction on the optical force:} The optical force on an achiral sphere with radius $a$ and dielectric constant $\epsilon_\mathrm{p}$ in an evanescent field with wavenumber $k=n_{2}\frac{\omega}{c}$ as a function of $ka$ using the static polarizability (red dashed) and the dynamic polarizability after Draine's correction (green continuous). Figures a) and b) correspond to the force on a dielectric sphere with $\epsilon_\mathrm{p}=2.56$ and an incident polarization corresponding to $\xi=1$ (linear diagonal polarization) respectively along the direction z (parallel to $\mathbf{p}^{o}$, figure a)) and the direction y (lateral optical force, figure b)). Figures c) and d) correspond to the lateral optical force using a gold sphere with $\epsilon_{\mathrm{p}}=-12.2+3i$ with an incident polarisation of the beam corresponding respectively  to $\xi=1$ (linear diagonal polarization, figure c)) and $\tau=1$ (linear horizontal polarisation, figure d)).\\

The force is normalized by $F0=\frac{a^{2}}{4\pi}I_{\mathrm{inc}}$, where $I_{\mathrm{inc}} = |\mathbf{E_{\mathrm{inc}}}|^2$ is the square modulus of the incident electric field. 
 The sphere suspended in water (index $n_{2}=1.33$) and the evanescent field is generated by total internal reflection in a prism of index $n_{1}=1.74$.Taking an incident intensity of $1.0\mathrm{mW}/\mathrm{\mu m}^{2}$ we have $F0/a^{2}=\frac{I_{\mathrm{inc}}}{4\pi}\simeq 3.3\cdot 10^{-4} \mathrm{N}.\mathrm{cm}^{-2}$. Here the incident angle of total internal reflection is $\theta\simeq 51\deg$ (so above the critical angle which is around $\theta=49,5\deg $). Note that in b) the force without correction is absolutely zero.}

\label{Comparison}
\end{figure}
\clearpage

\section{Evanescent field} 

\subsection{Snell-Descartes's Law}

When a propagating wave in a medium of index $n_{1}$ arrives at the interface with a medium $n$ with an incident angle $\theta$ the angle $\theta_{t}$ of the transmitted beam is given by Snell-Descartes's Law as:
\begin{equation}
\label{Snell}
n_{2}\sin(\theta_{\mathrm{t}})=n_{\mathrm{1}}sin(\theta)
\end{equation}

The amplitude of the transmitted wave can be expressed with respect to the incident polarization. If the incident electric field is oriented perpendicularly to the plane of incidence the wave is called s-polarized (or TE polarized). If the incident electric field is oriented parallely to the plane of incidence the wave is called p-polarized (or TM polarized). The transmission coefficient giving the amplitude of the transmitted wave for an incident s-wave and an incident a p-wave are given respectively by:
\begin{eqnarray}
\label{ts}
ts=\left| \frac{\mathbf{E_{\mathrm{t}}}}{\mathbf{E_{\mathrm{inc}}}}\right| _{s-wave}=\frac{2n_{\mathrm{1}}\cos(\theta)}{n_{\mathrm{1}}\cos(\theta)+n_{2}\cos(\theta_{\mathrm{t}})}\\
\label{tp}
tp=\left| \frac{\mathbf{E_{\mathrm{t}}}}{\mathbf{E_{\mathrm{inc}}}}\right| _{p-wave}=\frac{2n_{\mathrm{1}}\cos(\theta)}{n_{2}\cos(\theta)+n_{1}cos(\theta_{\mathrm{t}})}
\end{eqnarray}

Where $\mathbf{E_{\mathrm{t}}}$ stands for the transmitted field and $\mathbf{E_{\mathrm{inc}}}$ stands for the incident field in both case.\\
\\
Considering equation (\ref{Snell}) in case of $n_{2}<n_{\mathrm{1}}$ there exist a critical angle of the incidence $\theta_{c}$ such that  $\theta>\theta_{\mathrm{c}}$ implies $\sin(\theta_{\mathrm{t}})>1$. This correspond to a complex angle of transmission $\theta_{\mathrm{t}}$. In this regime the incident beam is totally internally reflected, giving rise to an evanescent field.\\
\\

\subsection{Structure of the EM field}

We can write the incident wave as:
\\
\begin{equation}
\label{incident}
\mathbf{E_{\mathrm{inc}}}=\frac{\sqrt{\mu_{\mathrm{1}}}A_{\mathrm{inc}}}{\sqrt{1+\arrowvert \eta_{\mathrm{inc}} \arrowvert^{\mathrm{2}}}}(\mathbf{e_{\mathrm{p}}}+\eta_{\mathrm{inc}}\mathbf{e_{\mathrm{s}}})\exp(i\mathbf{k_{\mathrm{inc}}}.\mathbf{r})
\end{equation}

Here $\mathbf{e_{\mathrm{s}}}$ and $\mathbf{e_{\mathrm{p}}}$ are the unit vector corresponding respectively to $s$ and $p$ polarization respectively. $A_{inc}$ correspond to the amplitude of the wave such that $\mu_{\mathrm{1}}\arrowvert A_{\mathrm{inc}}\arrowvert^{\mathrm{2}}=\Arrowvert\mathbf{E_{\mathrm{inc}}}\Arrowvert^{\mathrm{2}}$. $\Arrowvert k_{\mathrm{inc}}\Arrowvert=\frac{n_{\mathrm{1}}\omega}{\mathrm{c}}$ where $n_{\mathrm{1}}$ is the index of the medium as defined previously, $\mu_{\mathrm{1}}$ the magnetic permittivity of this medium, and $\epsilon_{\mathrm{1}}$ its permeability such that $n_{\mathrm{1}}=\sqrt{\epsilon_{\mathrm{1}}\mu_{\mathrm{1}}}$. A time dependency of $\exp(-i\omega t)$ is assumed.
It is also obvious in expression (\ref{incident}) that $\eta_{\mathrm{inc}}$ characterize the polarization, such that if $\eta_{\mathrm{inc}}=\pm i$ the wave is circularly polarized. In the same way $\eta_{\mathrm{inc}}=0$ is a purely p polarized wave and $\eta_{inc}=\pm\infty$ is a purely s polarized wave. 
\\
Then after transmission through the interface when $\theta>\theta_{\mathrm{c}}$ we have the evanescent field:

\begin{equation}
\mathbf{E}=\sqrt{\mu}\mathbf{E_0}\mathrm{exp}(-\kappa x)\mathrm{exp}(ik_\mathrm{z}z)
 =\frac{\sqrt{\mu} A_{\mathrm{2}}}{\sqrt{1+\arrowvert \eta_{\mathrm{2}} \arrowvert^{2}}}\left( \begin{array}
 d 1 \\
 \frac{\eta_{\mathrm{2}}}{\alpha}\\
 -i\frac{\kappa}{k\alpha}\\
 \end{array}\right) \exp(-\kappa x)\exp(ik_{z} z)
\end{equation}

Here we defined $\alpha=\sin(\theta_{\mathrm{t}})=\frac{k_{\mathrm{z}}}{k}$ and $k=n_{2}\frac{\omega}{c}$ and $n_{2}$ is the medium above the prism where is the evanescent wave. As previously $A_{\mathrm{2}}$ correspond to the amplitude of the wave and can be derived with $A_{\mathrm{2}}=\alpha\sqrt{\frac{\mu_{\mathrm{1}}}{\mu}}\frac{\sqrt{\left| tp\right| ^{2}+\left| \eta_{\mathrm{inc}}ts\right| ^{2}}}{\sqrt{1+\left| \eta_{\mathrm{inc}}\right| ^{2}}}A_{\mathrm{inc}}$ and we define the intensity as $I_{\mathrm{e}}=\frac{|\mathbf{E_{0}}|^{2}}{1+\tau(\kappa/k_{\mathrm{z}})^{2}}\exp(-2\kappa x)=\arrowvert A_{\mathrm{2}}\arrowvert^{2}e^{(-2\kappa x)}$. Also $\mu$ the magnetic permittivity of this medium, and $\epsilon$ its permeability such that $n_{2}=\sqrt{\epsilon\mu}$. Also $\kappa=k\sqrt{\alpha^{2}-1}=-ik_{\mathrm{x}}$ as found in the main text. The polarisation is described by $\eta_{\mathrm{2}}=\eta_{\mathrm{inc}}\frac{ts}{tp}$. Equations (\ref{ts}) and (\ref{tp}) give in this case:
\begin{eqnarray}
ts=\frac{2n_{\mathrm{1}}\cos(\theta)}{n_{\mathrm{1}}\cos(\theta)+in_{2}\sqrt{\alpha^{\mathrm{2}}-1}}\\
tp=\frac{2n_{\mathrm{1}}\cos(\theta)}{n_{2}\cos(\theta)+in_{\mathrm{1}}\sqrt{\alpha^{\mathrm{2}}-1}}
\end{eqnarray}
Defining $\mathbf{E_{0}}=\frac{A_{\mathrm{2}}}{\sqrt{1+\arrowvert \eta_{\mathrm{2}} \arrowvert^{2}}}\left( \begin{array}
d 1 \\
\frac{\eta_{\mathrm{2}}}{\alpha}\\
-i\frac{\kappa}{k\alpha}\\
\end{array}\right)$
We obtain the form that appears in the main text.
\\
With this notations we can define the degree of linear, diagonal and circular polarization in the \{x,y\} plane as in the main text with:
\begin{eqnarray}
\sigma = \frac{-2\mathrm{Im}\left\{\left(\mathbf{E}\cdot\mathbf{\hat{x}}\right)\left(\mathbf{E^*}\cdot\mathbf{\hat{y}}\right)\right\}}{\left(|\mathbf{E}\cdot\mathbf{\hat{x}}|^2+|\mathbf{E}\cdot\mathbf{\hat{y}}|^2\right)}=\frac{2Im(\eta_{\mathrm{2}})}{1+\arrowvert \eta_{\mathrm{2}}\arrowvert^{2}}\\ \xi=\frac{2\mathrm{Re}\left\{\left(\mathbf{E}\cdot\mathbf{\hat{x}}\right)\left(\mathbf{E^*}\cdot\mathbf{\hat{y}}\right)\right\}}{\left(|\mathbf{E}\cdot\mathbf{\hat{x}}|^2+|\mathbf{E}\cdot\mathbf{\hat{y}}|^2\right)}=\frac{2Re(\eta_{\mathrm{2}})}{1+\arrowvert \eta_{\mathrm{2}} \arrowvert^{2}}\\
\tau=\frac{(|\mathbf{E}\cdot\mathbf{\hat{x}}|^{2}-|\mathbf{E}\cdot\mathbf{\hat{y}}|^{2})}{\left(|\mathbf{E}\cdot\mathbf{\hat{x}}|^2+|\mathbf{E}\cdot\mathbf{\hat{y}}|^2\right)}=\frac{1-\arrowvert \eta_{\mathrm{2}} \arrowvert^{2}}{1+\arrowvert \eta_{\mathrm{2}} \arrowvert^{2}}
\end{eqnarray}
So in terms of the parameter $\eta_{inc}$ which define the incident polarisation:
\begin{equation}
\sigma=\frac{2Im(\eta_{\mathrm{inc}}\frac{ts}{tp})}{1+\arrowvert \eta_{\mathrm{inc}}\frac{ts}{tp} \arrowvert^{2}},   \xi=\frac{2Re(\eta_{\mathrm{inc}}\frac{ts}{tp})}{1+\arrowvert \eta_{\mathrm{inc}}\frac{ts}{tp} \arrowvert^{2}},  \tau=\frac{1-\arrowvert \eta_{\mathrm{inc}}\frac{ts}{tp} \arrowvert^{2}}{1+\arrowvert \eta_{\mathrm{inc}}\frac{ts}{tp} \arrowvert^{2}}
\end{equation}
These parameters are defined such that $\tau=1$ corresponds to a p-wave, $\tau=-1$ corresponds to a $s-wave$, $\sigma=1$ to a left circularly polarized plane wave in the x-y plane, $\sigma=-1$ to a right circularly polarized wave in the x-y plane, and $\xi=\pm 1$ correspond to opposite diagonal polarizations in the x-y plane.

\section{Derivation of the equations (5) and (6)}

\subsection{Derivation of $\mathbf{F_{int}}$}:
Considering equations (1)-(3) in the main text we have:
\begin{equation}
\begin{split}
\mathbf{F_{int}} = 		-\frac{k^4}{3}\mathrm{Re}\{ \sqrt{\frac{1}{\mu\epsilon}}\left[\alpha_\mathrm{e}\alpha_\mathrm{m}^* \left(\mathbf{E}\times\mathbf{H}^*\right) 	\right. \\
+i\alpha_\mathrm{e}\chi^*\left(\mathbf{E}\times\mathbf{E}^*\right)\\
+i\alpha_\mathrm{m}^*\chi\left(\mathbf{H}\times\mathbf{H}^*\right)\\
+|\chi|^2\left(\mathbf{E^{*}}\times\mathbf{H}\right)]\},
\end{split}
\label{force3}
\end{equation}

Using that $\mathrm{Re}\left\lbrace \mathbf{E}\times\mathbf{H^{*}}\right\rbrace =\frac{2c}{g}\mathbf{p}=\frac{2\omega}{g}\mathbf{p}\frac{\sqrt{\mu\epsilon}}{k}$ and $\mathrm{Im}\{\mathbf{E}\times\mathbf{E^{*}}\}=\frac{-4\omega\mu}{g}\mathbf{s_{\mathrm{e}}}$ and $\mathrm{Im}\{\mathbf{H}\times\mathbf{H^{*}}\}=\frac{-4\omega\epsilon}{g}\mathbf{s_{\mathrm{m}}}$ and $\mathrm{Re}\{\mathbf{E}\times\mathbf{E^{*}}\}=\mathrm{Re}\{\mathbf{H}\times\mathbf{H^{*}}\}=0$ and using $p''=\frac{g}{2c}\mathrm{Im}\{\mathbf{E}\times\mathrm{H^{*}}\}$ we obtain:
\begin{equation}
\label{f2}
\mathbf{F_{\mathrm{int}}}=-\frac{2\omega}{g}\frac{k^{3}}{3}(\mathrm{Re}\left\{\alpha_{\mathrm{e}} \alpha_{\mathrm{m}}^{*}\right\}\mathbf{p}-\mathrm{Im}\left\{\alpha_{\mathrm{e}} \alpha_{\mathrm{m}}^{*}\right\}\mathbf{{p''}}+\arrowvert\chi\arrowvert^{2}\mathbf{p})-\frac{2\omega}{g}\frac{2k^{4}}{3n_{2}}( \epsilon\mathrm{Re}\left\{\chi\alpha_{\mathrm{m}}^{*}\right\}\mathbf{s_{\mathrm{m}}}+ \mu\mathrm{Re}\left\{\chi\alpha_{\mathrm{e}}^{*}\right\}\mathbf{s_{\mathrm{e}}})
\end{equation}
As it appears in the main text in equation (6)
\\

\subsection{Derivation of $\mathbf{F_{\mathrm{0}}}$}
\\
From equation (3) in the main text we have: 
\begin{equation}
\begin{split}\mathbf{F_0} = \frac{1}{2}\mathrm{Re}\{	\alpha_e \mathbf{E}\left(\nabla\otimes\mathbf{E^*}\right) + \alpha_m \mathbf{H}\left(\nabla\otimes\mathbf{H^*}\right)\\
+ i\chi\left[\mathbf{H}\left(\nabla\otimes\mathbf{E^*}\right) - \mathbf{E}\left(\nabla\otimes\mathbf{H^*}\right)\right]\}\end{split}
\label{force2}
\end{equation}
Using $\mathbf{W}\left(\nabla\otimes\mathbf{V}\right)=\left(\mathbf{W}\cdot\nabla\right)\mathbf{V} + \mathbf{W}\times\left(\nabla\times\mathbf{V}\right)$ and the definitions of the table above
we can write the first term as \cite{Nieto2}:
\begin{equation}
\frac{1}{2}\mathrm{Re}\{	\alpha_\mathrm{e} \mathbf{E}\left(\nabla\otimes\mathbf{E^*}\right)\}=\frac{2\omega}{g}\mu\mathrm{Im}\{\alpha_{\mathrm{e}}\}\mathbf{p^{o}_{\mathrm{e}}}+\frac{1}{2}\mathrm{Re}\{\alpha_{e}\}\frac{1}{2}\nabla|\mathbf{E}|^{2}
\end{equation}
with $\mathbf{p^{o}}$ defined previously in the table in supplementary figure(1). Using that $\frac{1}{2}\nabla\arrowvert\mathbf{E}\arrowvert^{2}=\frac{2}{\epsilon g}\nabla u_{\mathrm_{e}}$. Similarly rewriting the magnetic term, it follows that:
\begin{equation}
\frac{1}{2}\mathrm{Re}\{	\alpha_e \mathbf{E}\left(\nabla\otimes\mathbf{E^*}\right) + \alpha_m \mathbf{H}\left(\nabla\otimes\mathbf{H^*}\right)\}=\frac{1}{g}\left( \epsilon^{-1}\mathrm{Re}\left\{\alpha_{\mathrm{e}}\right\}\nabla u_{\mathrm{e}}+  \mu^{-1} \mathrm{Re}\left\{\alpha_{\mathrm{m}}\right\}\nabla u_{\mathrm{m}}\right)+\frac{2\omega}{g}(\mu\mathrm{Im}\left\{\alpha_{\mathrm{e}}\right\}\mathbf{p_{\mathrm{e}}^{o}}+\epsilon\mathrm{Im}\left\{\alpha_{\mathrm{m}}\right\}\mathbf{p_{\mathrm{m}}^{o}})
\end{equation}
This is the achiral part of the force.
Now considering the third and fourth term and using again the identity $\mathbf{W}\left(\nabla\otimes\mathbf{V}\right)=\left(\mathbf{W}\cdot\nabla\right)\mathbf{V} + \mathbf{W}\times\left(\nabla\times\mathbf{V}\right)$ we obtain:
\begin{equation}
\begin{split}
\frac{1}{2}\mathrm{Re}\{i\chi\left[\mathbf{H}\left(\nabla\otimes\mathbf{E^*}\right) - \mathbf{E}\left(\nabla\otimes\mathbf{H^*}\right)\right]\}=\frac{1}{2}\mathrm{Re}\{i\chi\left[(\mathbf{H}\cdot\nabla)\mathbf{E^{*}}-(\mathbf{E}\cdot\nabla)\mathbf{H^{*}}+\mathbf{H}\times(\nabla\times \mathbf{E^{*}})-\mathbf{E}\times(\nabla\times\mathbf{H^{*}})\right]\}\\
=-\frac{1}{2}\mathrm{Im}\{\chi\left[(\mathbf{H}\cdot\nabla)\mathbf{E^{*}}-(\mathbf{E}\cdot\nabla)\mathbf{H^{*}}+\mathbf{H}\times(\nabla\times \mathbf{E}^{*})-\mathbf{E}\times(\nabla\times\mathbf{H^{*}})\right]\}\\
=-\frac{1}{2}\mathrm{Re}\{\chi\}\mathrm{Im}\{\nabla(\mathbf{E^{*}}\cdot\mathbf{H})\}-\frac{1}{2}\mathrm{Im}\{\chi\}\mathrm{Re}\{(\mathbf{H}\cdot\nabla)\mathbf{E^{*}}-(\mathbf{E}\cdot\nabla)\mathbf{H^{*}}\}\\-\frac{1}{2}\mathrm{Im}\{\chi\}\mathrm{Re}\{\mathbf{H}\times(\nabla\times \mathbf{E^{*}})-\mathbf{E}\times(\nabla\times\mathbf{H^{*}})\}-\frac{1}{2}\mathrm{Re}\{\chi\}\mathrm{Im}\{\mathbf{H}\times(\nabla\times \mathbf{E^{*}})-\mathbf{E}\times(\nabla\times\mathbf{H^{*}})\}
\end{split}
\end{equation}
Then using Maxwell's equations in gaussian units \cite{bliokh2014} $\nabla\times\mathbf{E^{*}}=-in_{2}k\epsilon^{-1}\mathbf{H^{*}}$ and $\nabla\times\mathbf{H^{*}}=ik n_{2}\mu^{-1}\mathbf{E^{*}}$
and the definition of helicity: $h=\frac{g}{2\omega}\mathrm{Im}\{\mathbf{E}^{*}\cdot\mathbf{H}\}$
we obtain:
\begin{equation}
\begin{split}
\frac{1}{2}\mathrm{Re}\{i\chi\left[\mathbf{H}\left(\nabla\otimes\mathbf{E^*}\right) - \mathbf{E}\left(\nabla\otimes\mathbf{H^*}\right)\right]\}=-\frac{\omega}{g}\mathrm{Re}\{\chi\}\nabla h-\frac{1}{2}\mathrm{Im}\{\chi\}\nabla\times\mathrm{Re}\{\mathbf{E}\times\mathbf{H^{*}}\}\\
+\frac{1}{2}\mathrm{Im}\{\chi\}\mathrm{Re}\{ik n_{2}[\epsilon^{-1}\mathbf{H}\times(\mathbf{H^{*}})+\mu^{-1}\mathbf{E}\times(\mathbf{E^{*}})]\}-\frac{1}{2}\mathrm{Re}\{\chi\}\mathrm{Im}\{ik n_{2}[\epsilon^{-1}\mathbf{H}\times(\mathbf{H^{*}})+\mu^{-1}\mathbf{E}\times(\mathbf{E^{*}})]\}
\end{split}
\end{equation}
And using that $\frac{2c}{g}\mathbf{p}=\mathrm{Re}\{\mathbf{E}\times\mathbf{H}^{*}\}$ and  $\mathrm{Re}\{[\epsilon^{-1}\mathbf{H}\times(\mathbf{H^{*}})+\mu^{-1}\mathbf{E}\times(\mathbf{E^{*}})]\}=0$ we have:
\begin{equation}
\frac{1}{2}\mathrm{Re}\{i\chi\left[\mathbf{H}\left(\nabla\otimes\mathbf{E^*}\right) - \mathbf{E}\left(\nabla\otimes\mathbf{H^*}\right)\right]\}=-\frac{\omega}{g}\mathrm{Re}\{\chi\}\nabla h -\frac{c}{g}\mathrm{Im}\{\chi\}\nabla\times\mathbf{p}-\frac{1}{2}\mathrm{Im}\{\chi\}\mathrm{Im}\{kn_{2}[\epsilon^{-1}\mathbf{H}\times\mathbf{H^{*}}+\mu^{-1}\mathbf{E}\times\mathbf{E^{*}}]\}
\end{equation} 
Finally using that $n_{2}\mathrm{Im}\{\epsilon^{-1}\mathbf{H}\times(\mathbf{H^{*}})+\mu^{-1}\mathbf{E}\times(\mathbf{E^{*}})\}=-\frac{4n_{2}\omega}{g}\mathbf{s}=-\frac{4c}{g}k\mathbf{s}$ we obtain:
\begin{equation}
\frac{1}{2}\mathrm{Re}\{i\chi\left[\mathbf{H}\left(\nabla\otimes\mathbf{E^*}\right) - \mathbf{E}\left(\nabla\otimes\mathbf{H^*}\right)\right]\}=-\frac{\omega}{g}\mathrm{Re}\{\chi\}\nabla h -\frac{c}{g}\mathrm{Im}\{\chi\}(\nabla\times\mathbf{p}-2k^{2}\mathbf{s})
\end{equation} 
This is the chiral part of the force.
So combining the four terms of $\mathbf{F_{\mathrm{0}}}$ we finally have:
\begin{equation}
\mathbf{F_{\mathrm{0}}}=\frac{1}{g}\left( \epsilon^{-1}\mathrm{Re}\left\{\alpha_{\mathrm{e}}\right\}\nabla u_{\mathrm{e}}+  \mu^{-1} \mathrm{Re}\left\{\alpha_{\mathrm{m}}\right\}\nabla u_{\mathrm{m}}\right)+\frac{2\omega}{g}(\mu\mathrm{Im}\left\{\alpha_{\mathrm{e}}\right\}\mathbf{p_{\mathrm{e}}^{o}}+\epsilon\mathrm{Im}\left\{\alpha_{\mathrm{m}}\right\}\mathbf{p_{\mathrm{m}}^{o}})-\frac{\omega}{g} \mathrm{Re}\left\{\chi\right\}\nabla h -\frac{c}{g}\mathrm{Im}\left\{\chi\right\}[\nabla\times\mathbf{p}-2k^{2}\mathbf{s}]
\end{equation}
As it appears in the main text in equation (5).

\section{Optical Forces on a molecule, a nano sphere and a gold nano helix}

In this section we briefly discuss the lateral forces that arise on three types of small chiral particles that commonly arise in the literature, and leave the details of the associated calculations to the next section.

\subsection{Helicene Molecules}

The optical manipulation of objects smaller than a few 100 nm in liquid suspension is in general very challenging due to thermal agitation. However, the optical sorting of individual atoms or molecules by chirality may nevertheless be possible in low-pressure environments or by tayloring the fields in modern metamaterials. A molecule of hexahelicene has a real chiral polarizability of $\chi = -6.2\cdot 10^{-2}\ \AA^3$ at a wavelength of 589.3 nm, which lies well outside of the molecule's absorption band.\cite{barron} At the same wavelength, the static electric polarizability of hexahelicene is $\alpha_{\mathrm{e}}=10.4\ \AA^3$ with the magnetic polarizability $\arrowvert \alpha_{\mathrm{m}}\arrowvert<10^{-5}|\alpha_{\mathrm{e}}|$. We consider a beam with a given intensity that is totally internally reflected at the interface of heavy flint glass ($n_1 = 1.74$) and air ($n_2 = 1$) at an angle of $\theta = 36.4^\circ$, which is close to the critical angle and gives the strongest force given the angle-dependent intensity and wave-vector of the evanescent field. Situated 60 nm above the interface, the hexahelicene molecule experiences a lateral force of $ F_{\mathrm{y}}\simeq 2.04\cdot 10^{-19} \mathrm{pN}/(\mathrm{mW}/\mathrm{\mu m})$ if $\sigma = 1$ (circularly polarized in the (x,y)-plane) and $ F_{\mathrm{y}}\simeq 4.04\cdot 10^{-19} \mathrm{pN}/(\mathrm{mW}/\mathrm{\mu m})$ if $\tau = 1$ (p-polarized). Stronger optical forces are possible within the absorption band, where the chiral polarizability is complex and highly dependent on the wavelength. Coronene is a close relative of hexahelicene which is achiral but has the same size and electric polarizability as helicene. The lateral force on this achiral molecule due to the Belifante momentum density $\mathbf{p^s}$ as predicted by Bliokh et. al. \cite{bliokh2014} is $F_{\mathrm{y}}\simeq 1.7.10^{-21} \mathrm{pN}$ for $\sigma =1$ and vanishes for $\sigma = 0$ (linearly polarized in (x,y) plane).

\subsection{Spherical Nanoparticles}

The polarizability of a chiral particle may be calculated using bi-isotropic constitutive relations for the material of the particle, where material chirality is parametrized with a chirality parameter $K\in [-1,1]$ \cite{wang2014}. The derivation and the somewhat bulky expressions for the polarizabilities of a small chiral sphere are given in the supplementary information. An example for spherical particles with highly chiral electromagnetic response are cholesteric liquid crystal droplets \cite{tkachenko2014}. Assuming the refractive index of 5CB (4-Cyano-4'-pentylbiphenyl) of $n_\mathrm{s}=1.597$ and $K=1$ for a 30 nm sphere in water with index $n_{2} = 1.33$, letting an incident beam with 589 nm wavelength totally internally reflect at $\theta = 52^\circ$ in heavy flint glass ($n_{1}=1.74$) causes a lateral force of $ F_{\mathrm{y}}\simeq 2.6\cdot10^{-5} \mathrm{pN}/(\mathrm{mW}/\mathrm{\mu m^2})$ if the sphere is floating 60 nm above the flint glass surface and $\tau = 1$.

The lateral force on a chiral sphere above a reflective surface in a propagating field as predicted by Wang et. al. \cite{wang2014} corresponds to $F_\mathrm{y} = 10^{-6}\mathrm{pN}/(\mathrm{mW/\mu m^2})$ for a sphere with 30nm radius in vacuum with a dielectric constant of $\epsilon = 2$ and chirality parameter $K=1$ when it is located 60 nm above a metal surface and illuminated with a plane wave that has $\tau = 1$ and a wavelength of 600 nm. In comparison, the same sphere would experience chirality-sorting lateral force of $ F_{\mathrm{y}}= 4.6\cdot 10^{-5}\mathrm{pN}/(\mathrm{mW/\mu m^{2}})$ at the same height above the surface in an evanescent field that is created by total internal reflection at $\theta = 36.4^\circ$ in heavy flint glass ($n_{1} = 1.74$).
\subsection{Gold Helix}

Increasing interest in chiral optical materials has recently driven the development of artificial nano structures with extreme chiral responses \cite{Gansel, gibbs, Rogacheva, McPeak}, including artificially chiral effective media \cite{yin2014}. Metallic nanohelices in particular enable the analytical estimation of their very strong chiral polarizability and have recently been fabricated at size scales small enough to reach plasmonic resonances in the visible range \cite{gibbs}. A perfectly conducting 50nm helix with 25nm diameter and one loop has a chiral polarizability of $\chi = 2.72\cdot10^{7}\ \AA^3$ and a static electric polarizability of $\alpha_\mathrm{e} = 2.56\cdot10^{8}\ \AA^3$ \cite{Jaggard}. Situated 30nm above the surface in an evanescent field with wavelength 589nm and $\tau =1$, a lateral force of $F_\mathrm{y} = -4.25\cdot10^{-3}\mathrm{pN}/(\mathrm{mW/\mu m^2})$ results if the beam is totally internally reflected at a flint glass / air interface at a $\theta = 36.4^\circ$ degree angle.

\section{Optical Forces on a sphere}
As stated in the main text if the sphere and the surrounding medium are assumed non magnetic the static polarisabilities are given at the first non-zero order by:
\begin{eqnarray}
\label{pole0}
\alpha_{\mathrm{e}}^{0}=\epsilon a^{3}\frac{(\epsilon_{\mathrm{p}}-\epsilon)}{\epsilon_{\mathrm{p}}+2\epsilon}\\
\label{polm0}
\alpha_{m}^{0}=\frac{\epsilon_{\mathrm{p}}-\epsilon}{30\epsilon}k^{2} a^{5}
\end{eqnarray}
Here $a$ refers to the radius of the sphere, $\epsilon_{\mathrm{p}}$ and $\epsilon$ are the relative permittivity of the sphere and the surrounding medium respectively. 
Using Draine's correction we obtain the dynamic polarizabilities as: $\alpha_{\mathrm{e}} = \left[1-\left(i2k^3/3\epsilon\right)\alpha_{\mathrm{e}}^{(0)} \right]^{-1}\alpha_{\mathrm{e}}^{(0)}$ and $\alpha_{\mathrm{m}} = \left[1-\left(i2k^3/3\right)\alpha_{\mathrm{m}}^{(0)} \right]^{-1}\alpha_{\mathrm{m}}^{(0)}$
The forces on an achiral sphere are then:
\begin{eqnarray}
\mathbf{F_{\mathrm{0}}}=-\frac{\kappa I_{\mathrm{e}}}{2}a^{3}\epsilon\mathrm{Re}\{\frac{\epsilon_{\mathrm{p}}-\epsilon}{\epsilon_{\mathrm{p}}+2\epsilon}\}(1+\tau\frac{\kappa^{2}}{k_{\mathrm{z}}^{2}})\mathbf{\hat{x}}\\
\frac{I_{\mathrm{e}}}{2}k_{\mathrm{z}}a^{3}(\epsilon\mathrm{Im}\{\frac{\epsilon_{\mathrm{p}}-\epsilon}{\epsilon_{\mathrm{p}}+2\epsilon}\}+\frac{2\epsilon}{3}(ka)^{3}\mathrm{Re}\{\left( \frac{\epsilon_{\mathrm{p}}-\epsilon}{\epsilon_{\mathrm{p}}+2\epsilon}\right)^{2} \})(1+\tau\frac{\kappa^{2}}{k_{\mathrm{z}}^{2}})\mathbf{\hat{z}}
\end{eqnarray}

\begin{eqnarray}
\mathbf{F_{\mathrm{int}}}=-I_{\mathrm{e}} \frac{k^{4}}{3} \frac{\kappa}{k_{z}}\arrowvert\frac{\epsilon_{\mathrm{p}}-\epsilon}{ 2\epsilon+\epsilon_{\mathrm{p}}}\arrowvert^{2}[\frac{1}{30}\mathrm{Re}\{\epsilon_{\mathrm{p}}+2\}k^{2}a^{8}\sigma +(\frac{1}{30}\mathrm{Im}\{\epsilon_{\mathrm{p}}\}k^{2} a^{8}-\frac{2}{90}k^{5}a^{11}\mathrm{Re}\{\frac{(\epsilon_{\mathrm{p}}-\epsilon)(\epsilon_{\mathrm{p}}+2\epsilon)^{*}}{\epsilon_{\mathrm{p}}+2\epsilon}\})\xi]\mathbf{\hat{y}}\\
-I_{\mathrm{e}}\frac{k^{4}}{3}\frac{k}{k_{\mathrm{z}}}[\frac{1}{30}\arrowvert\frac{\epsilon_{\mathrm{p}}-\epsilon}{ 2\epsilon+\epsilon_{\mathrm{p}}}\arrowvert^{2}\mathrm{Re}\{\epsilon_{\mathrm{p}}+2\}k^{2}a^{8}]\bf{\hat{z}}\\
-I_{\mathrm{e}}\frac{k^{4}}{3}\frac{\kappa k}{k_{\mathrm{z}}^{2}}\arrowvert\frac{\epsilon_{\mathrm{p}}-\epsilon}{ 2\epsilon+\epsilon_{\mathrm{p}}}\arrowvert^{2}[(\frac{1}{30}\mathrm{Im}\{\epsilon_{\mathrm{p}}\}k^{2} a^{8}-\frac{2}{90}k^{5}a^{11}\mathrm{Re}\{\frac{(\epsilon_{\mathrm{p}}-\epsilon)(\epsilon_{\mathrm{p}}+2\epsilon)^{*}}{\epsilon_{\mathrm{p}}+2\epsilon}\})\tau]\mathbf{\hat{x}}
\end{eqnarray}

For a chiral sphere the polarisabilities depend on the chirality of the sphere and of the surrounding medium. Assuming a non-chiral non magnetic surrounding medium and $\mu_{\mathrm{p}}=1$ Jaggard showed that we can derive:
\begin{eqnarray}
\alpha_{e}^{0}=a^{3}\epsilon\frac{[-2K^{2}\frac{\epsilon}{\epsilon_{\mathrm{p}}}+3(\epsilon{p}-\epsilon)]}{4 K^{2}\frac{\epsilon}{\epsilon_{p}}-3(2\epsilon+\epsilon_{p})}\\
\alpha_{m}^{0}=a^{3}\frac{[-2K^{2}\frac{\epsilon}{\epsilon_{p}}]}{4 K^{2}\frac{\epsilon}{\epsilon_{p}}-3(2\epsilon+\epsilon_{p})}\\
\chi=-3a^{3}\frac{K\epsilon}{4 K^{2}\frac{\epsilon}{\epsilon_{p}}-3(2\epsilon+\epsilon_{p})}
\end{eqnarray}
where $a$ is the radius of the sphere and K the chiral parameter such that the usual dispersion relation for two opposite circularly polarized plane wave in the medium of the sphere  is\cite{wang2014}: $k_{\pm}=(\sqrt{\epsilon_{\mathrm{p}}}\pm K)\frac{\omega}{c}$

\section{Chiral polarisability of an helix}
We know that we have the electric and magnetic dipole moments $\mathbf{d}$ and $\mathbf{m}$ induced by fields incident on a chiral particle with isotropic polarizabilities and no permanent dipole moment given by \cite{wang2014}:

\begin{equation}
\begin{array}{c}
\bf{d}=\alpha_{\mathrm{e}}\bf{E}+i\chi\bf{H} \\
\bf{m}=\alpha_{\mathrm{m}}\bf{H}-i\chi\bf{E}
\end{array}
\label{polarizabilities}
\end{equation}

Where $\alpha_{\mathrm{e}}$ and $\alpha_{\mathrm{m}}$ are the dynamic electric and magnetic polarizabilities. The chiral polarizability of the particle $\chi$ captures the chiral nature of the dipole, such that setting $\chi=0$ recovers the behavior for an achiral dipole. For a perfectly conducting helix in the Rayleigh regime and a perfect metal we know from \cite{Lakhtakia} that the polarisabilities are in gaussian units by:
\begin{eqnarray}
\chi=\alpha_{\mathrm{e}}(\frac{\pi a^{2}k}{2l})\\
\alpha_{\mathrm{e}}=\frac{(l)^{2}}{4\pi k^{2} a (L/\mu_{0}a) }
\end{eqnarray}
Where a is the radius of the helix, l its length taken and L its inductance in SI units and here $L/\mu_{0}a$ is taken in SI units. For an helix with one loop we have: $L\simeq \frac{\mu_{0}a^{2}\pi}{l}(1-\frac{16a}{3\pi l}+\frac{2a^{2}}{l^{2}}-\frac{4a^{4}}{l^{4}})(\mathrm{SI})$
we obtain:
\begin{eqnarray}
\chi\simeq\frac{l^{2}}{4\pi k}(1-\frac{16a}{3\pi l}+\frac{2a^{2}}{l^{2}}-\frac{4a^{4}}{l^{4}})^{-1}\\
\alpha_{\mathrm{e}}\simeq\frac{l^{3}}{4\pi^{2}k^{2}a^{2}}(1-\frac{16a}{3\pi l}+\frac{2a^{2}}{l^{2}}-\frac{4a^{4}}{l^{4}})^{-1}
\end{eqnarray}
Using l=50nm and a=12.5nm so that the diameter is 2a=l/2=25nm we obtain: $\frac{1}{1-\frac{16a}{3\pi l}+\frac{2a^{2}}{l^{2}}-\frac{4a^{4}}{l^{4}}}\simeq 1.46$

\section{Supplementary References}